\documentclass[apjl]{emulateapj} 

\usepackage{epsfig}
\usepackage{amsmath}
\usepackage{amssymb}
\usepackage{natbib}
\usepackage{threeparttable} 
\usepackage{booktabs}

\usepackage{epstopdf}

\newcommand{\chan}{\textit{Chandra}}
\newcommand{\swift}{\textit{Swift}}
\newcommand{\rxte}{\textit{RXTE}}

\newcommand{\inte}{\textit{INTEGRAL}}

\newcommand{\beppo}{\textit{BeppoSAX}}

\newcommand{\Msun}{\mathrm{M}_{\odot}}
\newcommand{\lum}{\mathrm{erg~s}^{-1}}
\newcommand{\flux}{\mathrm{erg~cm}^{-2}~\mathrm{s}^{-1}}
\newcommand{\fluence}{\mathrm{erg~cm}^{-2}}
\newcommand{\cnts}{\mathrm{c~s}^{-1}}

\newcommand{\nh}{\mathrm{cm}^{-2}}

\newcommand{\dist}{(D/5.0~\mathrm{kpc})^2}
\newcommand{\ioniz}{\mathrm{erg~cm~s}^{-1}}

\newcommand{\source}{IGR J17062--6143}

\newcommand{\gro}{GRO J1655--40}

\slugcomment{To appear in the Astrophysical Journal Letters}

\shorttitle{An energetic X-ray burst from IGR J17062--6143}
\shortauthors{Degenaar et al.}

\begin{document}


\title{X-ray emission and absorption features during an energetic thermonuclear X-ray burst from IGR J17062--6143}

\author{N. Degenaar$^{1,}$\altaffilmark{4}, J. M. Miller$^{1}$, R. Wijnands$^{2}$, D. Altamirano$^{2}$ and A. C. Fabian$^{3}$}
\affil{$^1$Department of Astronomy, University of Michigan, 500 Church Street, Ann Arbor, MI 48109, USA; degenaar@umich.edu\\
$^2$Astronomical Institute Anton Pannekoek, University of Amsterdam, Postbus 94249, 1090 GE Amsterdam, The Netherlands\\
$^3$Institute of Astronomy, University of Cambridge, Madingley Road, Cambridge CB3 OHA, UK\\}

\altaffiltext{4}{Hubble fellow}


\begin{abstract}
Type-I X-ray bursts are thermonuclear explosions occurring in the surface layers of accreting neutron stars. These events are powerful probes of the physics of neutron stars and their surrounding accretion flow. We analyze a very energetic type-I X-ray burst from the neutron star low-mass X-ray binary \source\ that was detected with \swift\ on 2012 June 25. The light curve of the $\simeq$18-min long X-ray burst tail shows an episode of $\simeq$10~min during which the intensity is strongly fluctuating by a factor of $\simeq$3 above and below the underlying decay trend, on a time scale of seconds. The X-ray spectrum reveals a highly significant emission line around $\simeq$1~keV, which can be interpreted as a Fe-L shell line caused by irradiation of cold gas. We also detect significant absorption lines and edges in the Fe-K band, which are strongly suggestive of the presence of hot, highly ionized gas along the line of sight. None of these features are present in the persistent X-ray spectrum of the source. The time scale of the strong intensity variations, the velocity width of the Fe-L emission line (assuming Keplerian motion), and photoionization modeling of the Fe-K absorption features each independently point to gas at a radius of $\simeq10^{3}$~km as the source of these features. The unusual X-ray light curve and spectral properties could have plausibly been caused by a disruption of the accretion disk due to the super-Eddington fluxes reached during the X-ray burst.
\end{abstract}



\section{Introduction}\label{sec:intro}
Type-I X-ray bursts are intense flashes of X-ray emission that have a duration of seconds to hours. These events are caused by unstable thermonuclear burning, which transforms the hydrogen and/or helium that is accreted onto the surface of a neutron star into heavier elements \citep[for reviews, see][]{lewin95,strohmayer06,schatz2006}. The light curves of X-ray bursts are characterized by a fast rise that is caused by burning of the fuel layer, followed by a slower decay phase that represents cooling of the burning ashes. Their X-ray spectra can be modeled by blackbody emission that peaks at a temperature of $\simeq$2--3~keV and cools to $\simeq$1~keV in the tail of the X-ray burst. 

The peak radiation of X-ray bursts can exceed the Eddington limit, which causes the photosphere to expand. This can be observed as an increase in black body emission radius (typically by a factor of a few) that is accompanied by a decrease in effective temperature while maintaining an approximately constant flux. On rare occasions, X-ray bursts have been observed to be so powerful that the emission radius increased by a factor of $\simeq$100. This is denoted as superexpansion \citep[][]{zand2010}.

X-ray bursts are a unique signature of neutron star low-mass X-ray binaries (LMXBs). In LMXBs, a Roche-lobe overflowing late-type companion star feeds matter to the compact object via an accretion disk. There is a delicate connection between the properties of X-ray bursts and that of the accretion flow. For example,  the rate at which mass is accreted onto the neutron star determines the duration and recurrence time of X-ray bursts \citep[e.g.,][]{fujimoto81,bildsten98,peng2007}, and the accretion geometry can strongly influence their observable properties \citep[e.g.,][]{cavecchi2011,linares2012_ter5,zand2012}. Vice versa, there are several lines of evidence suggesting that powerful X-ray bursts can affect the accretion flow \citep[][]{yu1999,strohmayer2002,ballantyne2004,kuulkers2008,chen2011,zand2011,altamirano2012}.


\source\ is an X-ray source that was discovered with \inte\ in 2006 \citep[][]{churazov2007}. It has been active ever since, displaying a 2--10 keV luminosity of $L_{\mathrm{X}}\simeq(1-5)\times10^{35}~\dist~\lum$ \citep[][]{ricci2008,remillard2008,degenaar2012_igrburster}. \source\ remained unclassified until \swift's Burst Alert Telescope \citep[BAT;][]{barthelmy05} detected an X-ray burst on 2012 June 25 \citep[][]{barthelmy2012}, which identified the source as an accreting neutron star LMXB \citep[][]{degenaar2012_igrburster}. In this Letter we report on unusual features present in the light curve and spectrum of this X-ray burst.


\section{Observations, Data Analysis and Results}\label{sec:data}
The BAT triggered on \source\ on UT 2012 June 25 at 22:42 \citep[][]{barthelmy2012}. The X-ray Telescope \citep[XRT;][]{burrows05} began to observe the source in the windowed timing (WT) mode $\simeq$157~s later, for a total exposure time of $927$~s (Obs ID 525148000). We reduced and analyzed the \swift\ data using the \textsc{heasoft} software package (ver. 6.11). 

The BAT and XRT data were processed using the \textsc{batgrbproduct} and \textsc{xrtpipeline} tools, respectively. We extracted BAT light curves and spectra with \textsc{batbinevt}, and used \textsc{XSelect} for the XRT data. To avoid pile-up, we used an annular extraction region with an inner (outer) radius of 2 (40) pixels when the XRT count rate exceeded 200$~\cnts$, and radii of 1 (40) pixels when the count rate was 100--200$~\cnts$ \citep[][]{romano2006}. For lower intensities we used a circular region with a radius of 40 pixels. 

The XRT spectral data were grouped into bins with a minimum of 20 photons and fitted between 0.5--10 keV using \textsc{XSpec} \citep[ver. 12.7;][]{xspec}. The BAT spectrum was fitted between 15--40 keV. We accounted for interstellar absorption by using the TBABS model \citep[][]{wilms2000}. Quoted errors represent $1\sigma$ confidence intervals. 


\subsection{The BAT X-Ray Burst Peak}\label{subsec:bat}
The source is detected with the BAT for $\simeq$160~s, peaking at $t\simeq$80~s after the trigger. The average spectrum, extracted from $t=0-160$~s, can be described by a black body model with a temperature of $kT_{\mathrm{bb}}=2.6\pm0.3$~keV and an emitting radius of $R_{\mathrm{bb}}=4.9^{+2.8}_{-1.7}$~km (for an assumed distance of 5.0 kpc; see below). By extrapolating the fit to the 0.01--100 keV energy range, we estimate an average bolometric flux of $F_{\mathrm{bol}}=4.8^{+2.8}_{-1.6}\times10^{-8}~\flux$. This gives a fluence for the X-ray burst peak of $f_{\mathrm{BAT}}\simeq7.7\times10^{-6}~\fluence$. 

The peak count rate observed with the BAT is a factor of $\simeq$2.6 higher than the average intensity in the 160-s interval, which suggests that the bolometric flux peaked at $\simeq~1.25\times10^{-7}~\flux$. Equating this to the empirical Eddington limit of X-ray bursts \citep[$3.8\times10^{38}~\lum$;][]{kuulkers2003}, places the source at a distance of $D\simeq5.0$~kpc \citep[see also][]{degenaar2012_igrburster}.

 \begin{figure}
 \begin{center}
	\includegraphics[width=7.5cm]{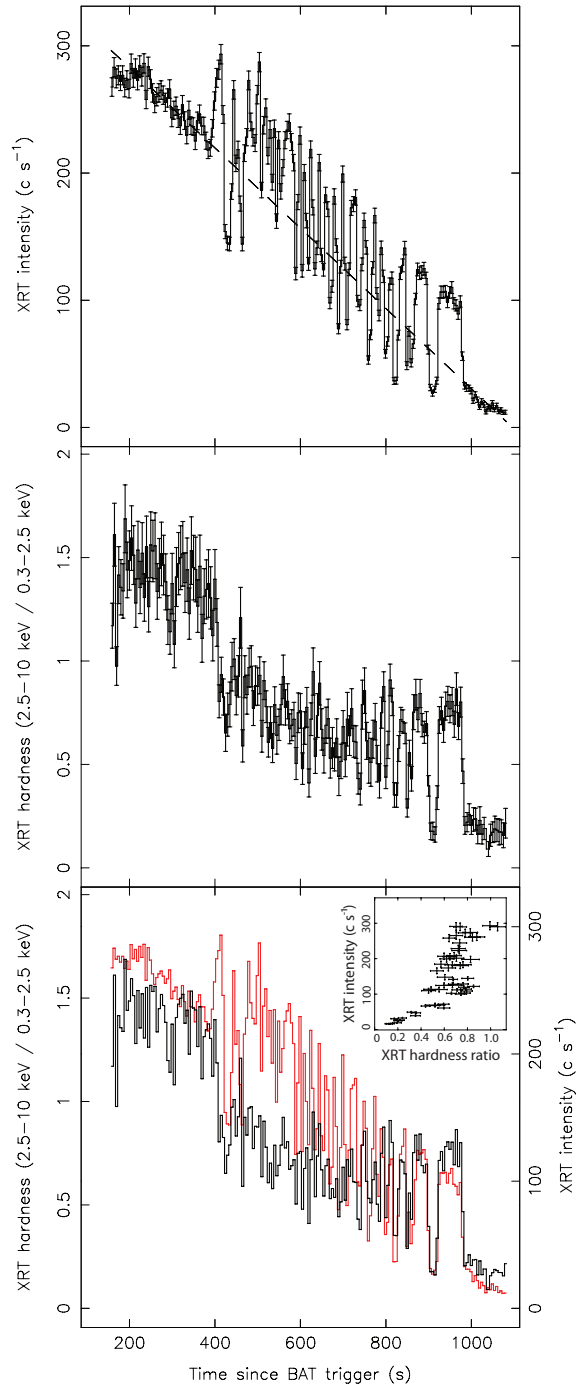}
    \end{center}
    \caption[]{\swift/XRT data of the X-ray burst tail at 5 s resolution. Top: count rate light curve (0.3--10 keV). The dashed line indicates a fit to a linear decay with a slope of $-$0.32~c~s$^{-2}$. Middle: ratio of counts in the 2.5--10 keV and 0.3--2.5 keV energy bands. Bottom: combined plot of the hardness ratio (black) and count rate light curve (red). Error bars were omitted for clarity. The inset displays the intensity versus the hardness during the fluctuating part of the light curve ($\simeq$390--980~s after the BAT trigger).
   }
 \label{fig:lc}
\end{figure}


\subsection{Light Curve of the X-Ray Burst Tail}\label{subsec:lc}
The XRT light curve shows a decrease in intensity from $\simeq$300 to $\simeq$10$~\cnts$ (Figure~\ref{fig:lc}). The overall decay is best described by a simple linear function with a slope of $-0.32$~c~s$^{-2}$ and a normalization of $296.5 \pm 0.9~\cnts$ at the start of the XRT observation. At $\simeq$390~s after the BAT trigger, the source intensity starts to strongly fluctuate by as much as a factor of $\simeq$3 above and below the underlying decay trend. The variations have a typical duration of seconds, and continue for $\simeq$590~s. Timing analysis did not reveal periodicities \citep[see also][]{degenaar2012_igrburster}. 

Investigation of the ratio of the counts in the 2.5--10 keV and 0.3--2.5 keV energy bands shows that the hardness decreased over time (i.e., the spectrum became softer; Figure~\ref{fig:lc}). The hardness ratio drops at the start of the fluctuation phase, but after that there is no apparent correlation between the intensity and the spectral hardness until $\simeq$250~s before the variations disappear. In that final part of the fluctuation phase the hardness ratio is positively correlated with the intensity (Figure~\ref{fig:lc}). The initial energy-independence suggests that the plasma causing the fluctuations is in thermal balance with the radiation field, which appears to break down shortly before the variations disappear.

 \begin{figure*}
 \begin{center}
	\includegraphics[width=8.8cm]{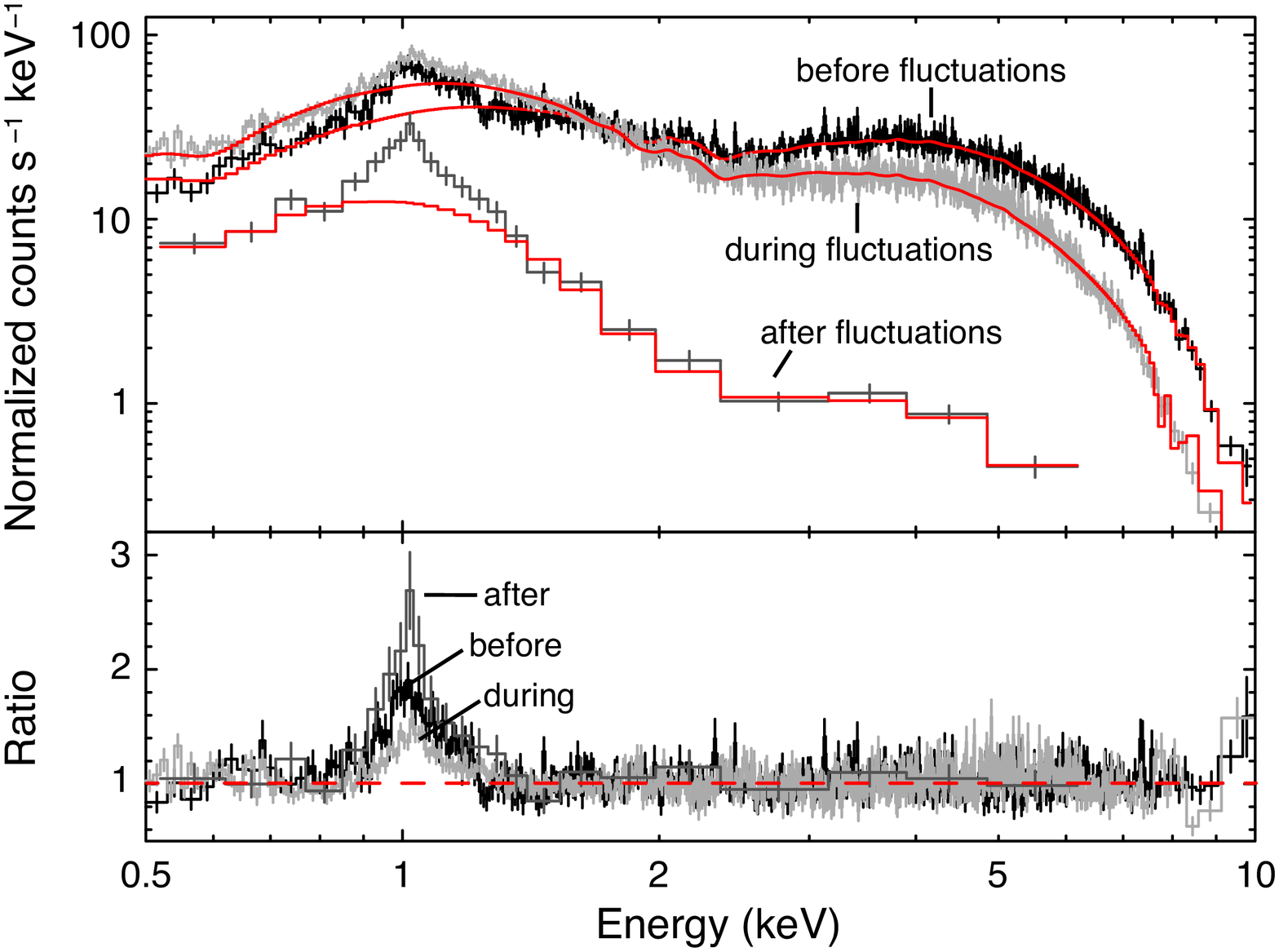}\hspace{0.1cm}
		\includegraphics[width=8.8cm]{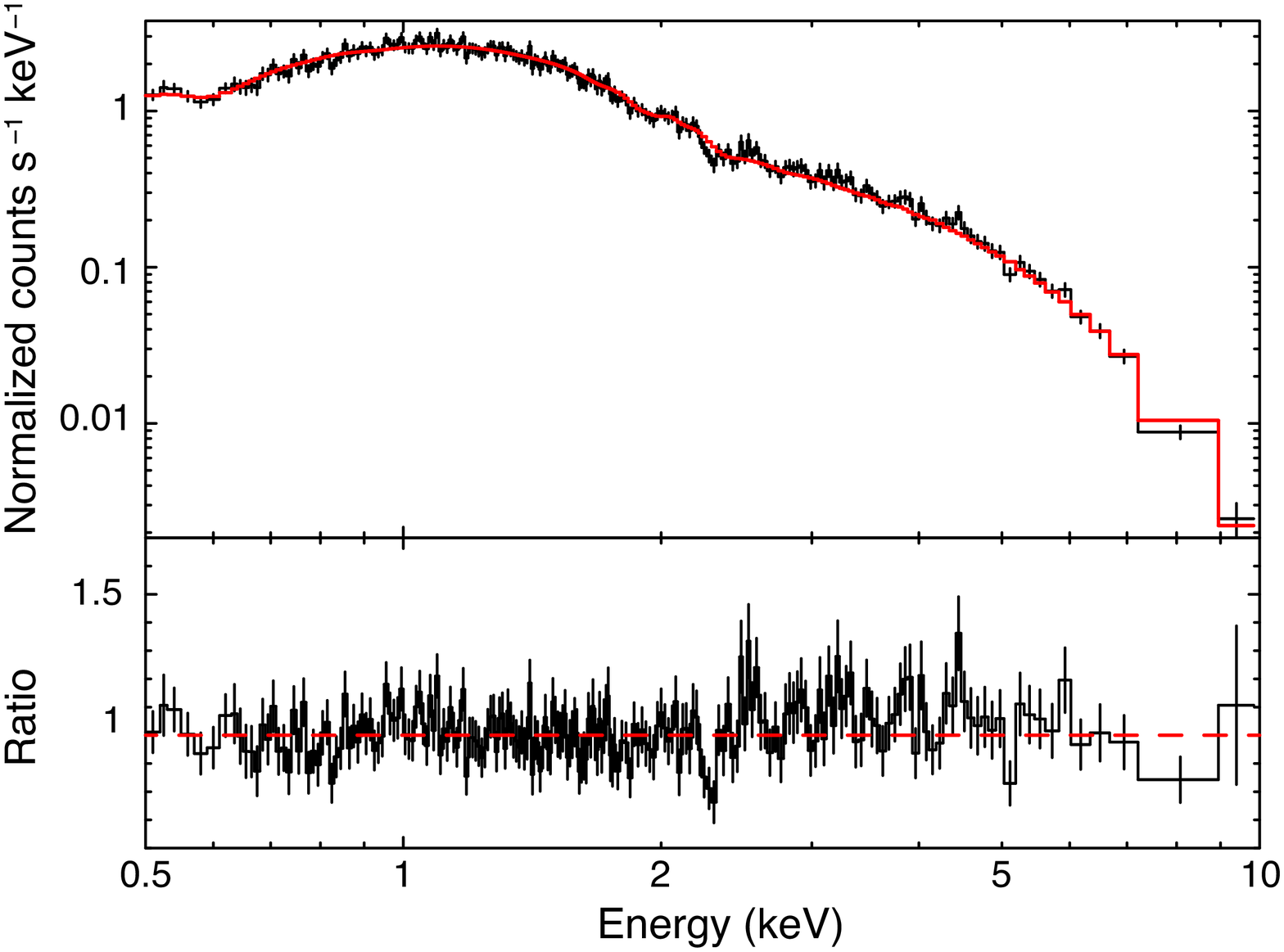}
    \caption[]{\swift/XRT WT mode spectra of \source\ (top panels) and data to model ratios (bottom panels). The solid red lines indicate the fitted continuum models. The spectra were rebinned for representation purposes. Left: X-ray burst spectra before (black), during (light grey), and after (dark grey) the fluctuation phase, with the normalization of the Fe-L emission line set to zero. Right: average spectrum of the persistent X-ray emission as observed in 2008 May \citep[][]{ricci2008,degenaar2012_igrburster}.
     \label{fig:spec}
    }
        \end{center}
\end{figure*}

 \begin{figure*}
 \begin{center}
	\includegraphics[width=8.8cm]{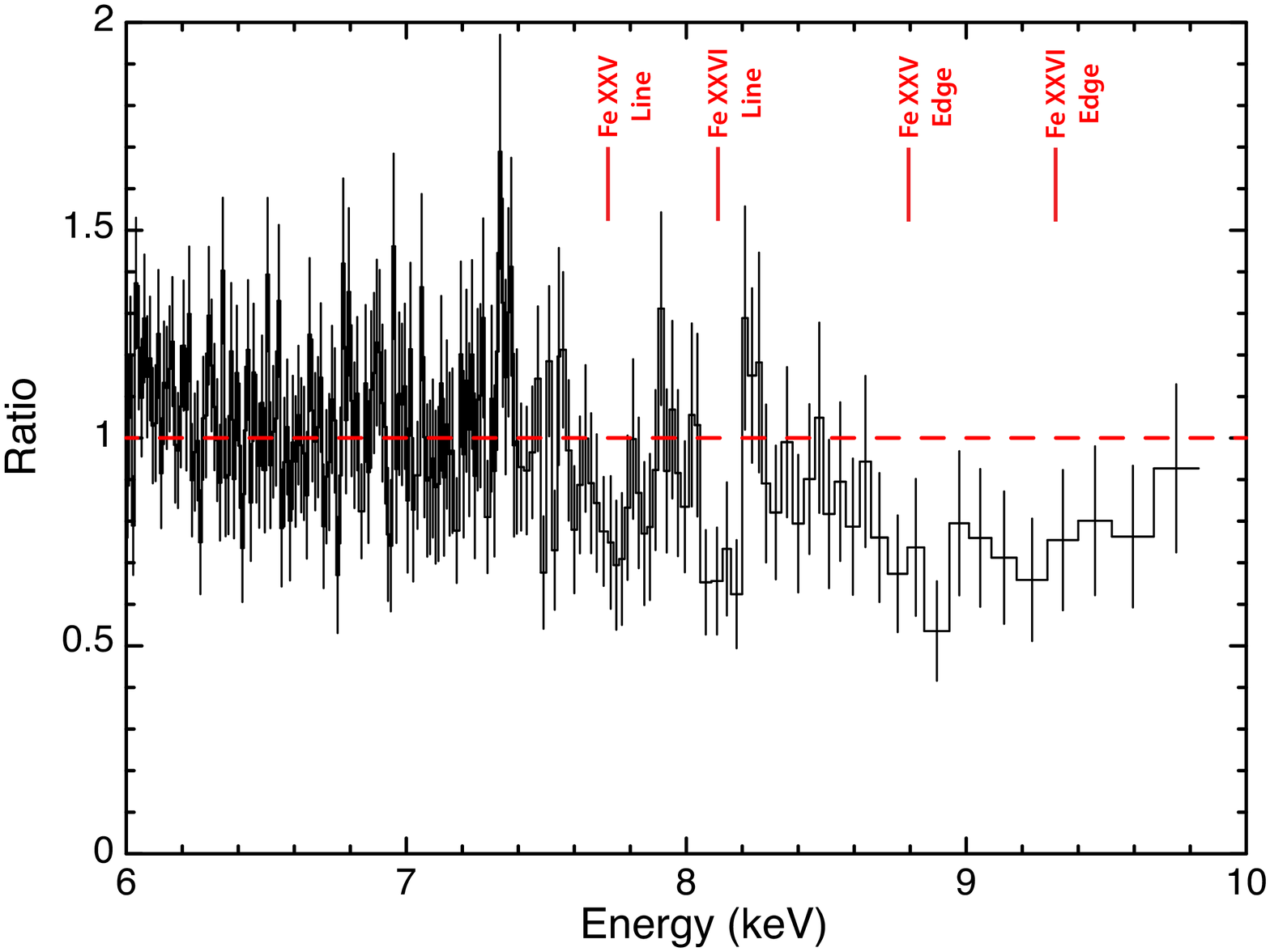}\hspace{0.1cm}
		\includegraphics[width=8.8cm]{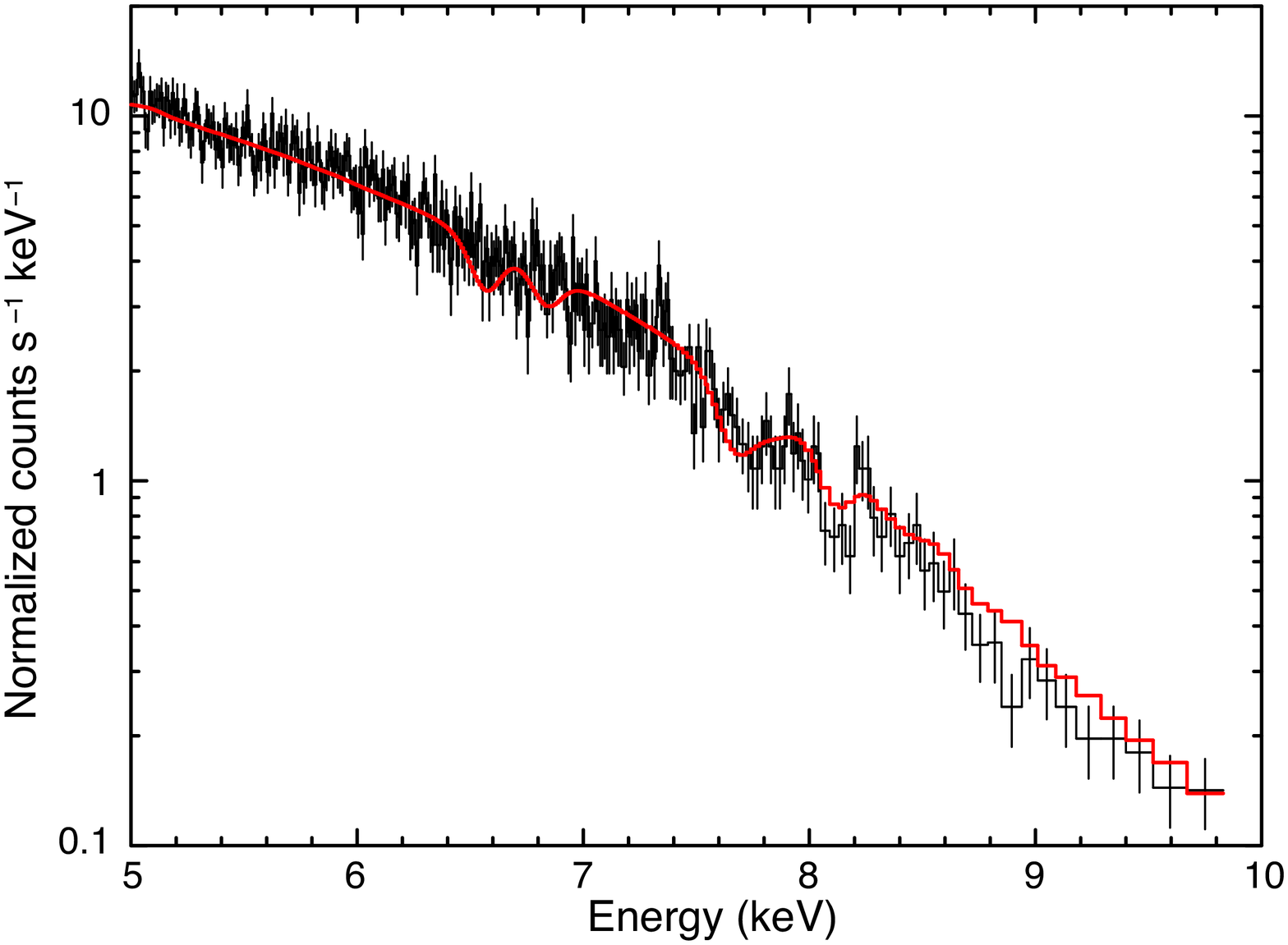}
    \caption[]{High-energy part of the average \swift/XRT spectrum of the X-ray burst tail. Left: Ratio of the data and the continuum model. Right: Spectral data compared to an \textsc{xstar} model developed for \gro\ \citep[solid red line; from][]{miller2008}.
     \label{fig:xstar}
    }
        \end{center}
\end{figure*}


~\\
\subsection{Spectrum of the X-Ray Burst Tail}\label{subsec:averagespec}
We investigate the average XRT spectrum to obtain a global characterization of the X-ray burst. A single absorbed black body does not provide an adequate description of the XRT data ($\chi_{\nu}^2>$2 for 744 dof). The fit can be improved by adding a power-law component ($\chi_{\nu}^2$=1.91 for 742 dof), or a second black body ($\chi_{\nu}^2$=1.72 for 742 dof), although neither provide a formally acceptable fit. Any reasonable combination of continuum components leaves a strong emission line at $\simeq$1~keV. 

Adding a Gaussian line to a continuum consisting of two black bodies improves the fit significantly ($\chi_{\nu}^2$=1.18 for 739 dof). The line has a centroid energy of $E_{\mathrm{l}}=1.018\pm0.004$~keV, a width of $\sigma = (6.73 \pm 0.44)\times10^{-2}$~keV, and a normalization of $(8.22 \pm 0.30)\times10^{-2}~\mathrm{photons~cm}^{-2}~\mathrm{s}^{-1}$ (Table~\ref{tab:spec}). These line properties correspond to an equivalent width of $105 \pm 3$~eV. Dividing the normalization by its error indicates that the line is highly significant at the $27\sigma$ level of confidence. Assuming that abundances are not unusual and velocity shifts are modest, the 1-keV feature is most likely an Fe L-shell line or possibly a combination of a few such lines from a range of charge states (Ne \textsc{x} is also in this range). The line is detected at all intervals during the burst (Section~\ref{subsubsec:timeresolved}), while it is not present in the persistent X-ray spectrum of the source (Figure~\ref{fig:spec}).

In addition to the 1-keV emission feature, there are structures visible in the Fe-K band between $\simeq$7--10 keV (Figure~\ref{fig:xstar}). Absorption lines consistent with Fe \textsc{xxv} He-$\beta$ (7.88 keV), Fe \textsc{xxvi} Ly-$\beta$ (8.25 keV), and K edges associated with Fe \textsc{xxv} (8.8 keV) and Fe \textsc{xxvi} (9.3 keV) are evident. Adding two edges and two Gaussian absorption lines further improves the spectral fit ($\chi_{\nu}^2$=1.07 for 733 dof). The normalizations of the Gaussians suggests that the absorption lines are significant at the 5$\sigma$--6$\sigma$ level of confidence. The inclusion of an edge at 8.8 keV is similarly significant (5.6$\sigma$), although the edge at 9.3 keV is statistically not required. The presence of these Fe-K band features is strongly suggestive of highly ionized absorption along the line of sight.  

The best-fit results for the average X-ray burst spectrum are listed in Table~\ref{tab:spec}. The temperature and emitting radius of the first black body component ($kT_{\mathrm{bb}}\simeq1.8$~keV and $R_{\mathrm{bb}}\simeq5$~km) are typical for the tails of X-ray bursts. The second black body is cooler and more extended ($kT_{\mathrm{bb}}\simeq0.3$~keV and $R_{\mathrm{bb}}\simeq80$~km), and could possibly arise from the (inner) accretion disk. The first black body (representing the X-ray burst emission) accounts for $\simeq$90\% of the total unabsorbed 0.5--10 keV flux of $F_{\mathrm{X}}=(1.16\pm0.01)\times10^{-8}~\flux$. By extrapolating this component to the 0.01--100 keV energy range while setting the normalization of all others to zero, we estimate a thermal bolometric flux of $F_{\mathrm{bol}}=(1.28\pm0.02) \times10^{-8}~\flux$ for the X-ray burst tail.

\begin{table*}
\begin{center}
\caption{Spectral Parameters for the X-Ray Burst Tail\label{tab:spec}}
\begin{tabular*}{0.98\textwidth}{@{\extracolsep{\fill}}llllll}
\hline
\hline
Model Component & Parameter (Unit) &  Average & Before Fluctuations & Fluctuations & After Fluctuations\\
\hline
TBABS  & $N_H~(10^{21}~\nh)$ & $1.63\pm0.04$ & \nodata & $1.43\pm0.14$ & \nodata \\	
BBODYRAD1  & $kT_{\mathrm{bb}}$~(keV) &   $1.787\pm0.005$ &   $2.082\pm0.025$ &   $1.584\pm0.011$  &   $1.423\pm0.013$\\  
BBODYRAD1 & $R_{\mathrm{bb}}$~(D/5.0 kpc km)  &  $5.49\pm0.01$ &  $6.22\pm0.03$ &  $5.43\pm0.02$ &  $1.53\pm0.16$  \\  
BBODYRAD2  & $kT_{\mathrm{bb}}$~(keV) &  $0.258 \pm 0.001$ &  $0.260 \pm 0.008$ &  $0.256 \pm 0.006$ &  $0.213 \pm 0.008$ \\  
BBODYRAD2 & $R_{\mathrm{bb}}$~(D/5.0 kpc km)  &  $83.56\pm0.01$ &  $81.33\pm0.18$ &  $86.60\pm0.14$ &  $68.18\pm0.20$ \\  	
GAUSSIAN1: Emission & $E_{\mathrm{l}}$ (keV) & $1.018\pm0.004$ & $1.017\pm0.006$ & $1.029\pm0.006$ & $1.000\pm0.010$ \\	
GAUSSIAN1: Emission & $\sigma$ ($10^{-2}$~keV) &  $6.73\pm0.44$ &  $7.76\pm0.84$ &  $7.52\pm1.28$ &  $7.16\pm1.97$ \\	
GAUSSIAN1: Emission & Norm. ($10^{-2}$~ph~cm$^{-2}$~s$^{-1}$) &  $8.22\pm0.30$ &  $15.2\pm1.25$ &  $6.69\pm0.68$ &  $4.64\pm0.73$ \\	
GAUSSIAN2: Abs. Fe \textsc{xxv}  & $E_{\mathrm{l}}$ (keV) &   $7.73\pm0.03$ &   \nodata &   $7.73$ fixed &   \nodata \\	
GAUSSIAN2: Abs. Fe \textsc{xxv}  & $\sigma$ (keV) &  0 fixed &  \nodata &  0 fixed & \nodata \\	
GAUSSIAN2: Abs. Fe \textsc{xxv}  & Norm. ($10^{-3}$~ph~cm$^{-2}$~s$^{-1}$) &  $5.9\pm1.2$ &  \nodata &  $4.3\pm0.8$ &  \nodata \\		
GAUSSIAN3: Abs. Fe \textsc{xxvi}  & $E_{\mathrm{l}}$ (keV) &   $8.10\pm0.03$ &   \nodata &   $8.10$ fixed &   \nodata\\	
GAUSSIAN3: Abs. Fe \textsc{xxvi}  & $\sigma$ (keV) &  0 fixed &  \nodata &  0 fixed &  \nodata \\	
GAUSSIAN3: Abs. Fe \textsc{xxvi}  & Norm. ($10^{-3}$~ph~cm$^{-2}$~s$^{-1}$) &  $6.6\pm1.2$ &  \nodata &  $4.4\pm0.9$ &  \nodata  \\	
EDGE: Fe \textsc{xxv}  & $E_{\mathrm{c}}$ (keV) &   8.8 fixed &   \nodata &   8.8 fixed &   \nodata \\	
EDGE: Fe \textsc{xxv}  & $\tau$ &  $0.50\pm0.09$ &  \nodata &  $0.40\pm0.12$ &  \nodata \\
EDGE: Fe \textsc{xxvi}  & $E_{\mathrm{c}}$ (keV) & 9.3 fix &  \nodata &  \nodata &  \nodata  \\	
EDGE: Fe \textsc{xxvi}  & $\tau$ &  $<0.07$ &  \nodata &  \nodata &  \nodata \\ 
\nodata & 	$\chi_{\nu}^2$/dof & 1.07/733 & \nodata & 1.17/1427 & \nodata \\
\hline
\end{tabular*}
\tablenotes{
{\bf Notes.} Quoted errors represent a $1\sigma$ confidence level. The spectra of the intervals before, during, and after the fluctuation phase were fitted simultaneously with $N_H$ tied.
}
\end{center}
\end{table*}


\subsubsection{Time Resolved Spectroscopy}\label{subsubsec:timeresolved}
We extracted separate spectra from intervals before, during, and after the fluctuation phase by selecting data from $t=157-390$, $391-980$ and $981-1084$~s since the BAT trigger, respectively. We fit the three spectra simultaneously using the best-fit model found from analyzing the average X-ray burst spectrum. The results are summarized in Table~\ref{tab:spec}.

The first black body displays the characteristic cooling seen during X-ray burst tails: the temperature decreases from $kT_{\mathrm{bb}}=2.08\pm 0.02$ to $1.42\pm 0.13$~keV (Table~\ref{tab:spec}). The fractional contribution of this model component to the total 0.5--10 keV unabsorbed flux decreases from $\simeq99\%$ to $\simeq68\%$. The 1-keV emission feature is detected during the entire X-ray burst with a significance of 12.2$\sigma$, 9.8$\sigma$, and 6.4$\sigma$ for the intervals before, during and after the fluctuation phase, respectively (Figure~\ref{fig:spec}). The line flux decreases over time, but the energy and width do not change significantly (Table~\ref{tab:spec}). The Fe-K band absorption lines are present during the fluctuation phase (4$\sigma$--5$\sigma$ level), as is the edge at 8.8 keV (3$\sigma$). These features are not significantly detected before and after this interval, which is plausibly due to the lower statistics. 


\subsubsection{X-Ray Spectra of the Peaks and Dips}\label{subsubsec:peakdipspec}
We extracted separate spectra for the the peaks and the dips of the fluctuation phase by separating the points lying above and below the linear decay fit (dashed line in Figure~\ref{fig:lc}). We fit the two spectra simultaneously using the best-fit model found from analyzing the average X-ray burst spectrum. 

The Fe-K band absorption features are not significantly detected in the separate spectra, but the Fe-L emission remains highly significant at the 8$\sigma$--10$\sigma$ level. We find that the line is slightly shifted towards the red for the dip spectrum when compared to that of the peaks; $E_{\mathrm{l,dips}}=1.028\pm0.007$ and $E_{\mathrm{l,peaks}}=1.050\pm0.010$~keV. The width of the line is similar in the two spectra ($\sigma_{\mathrm{peaks}}=$0.109$\pm$0.016~keV and $\sigma_{\mathrm{dips}}=$0.090$\pm$0.014~keV), but the normalization is larger for the dips; $(10.6\pm1.0)\times10^{-2}~\mathrm{photons~cm}^{-2}~\mathrm{s}^{-1}$, compared to $(7.5\pm1.1)\times10^{-2}~\mathrm{photons~cm}^{-2}~\mathrm{s}^{-1}$ for the peaks. These differences may indicate that emission from a lower charge state, at larger radius, is relatively more important during the dips.


\subsection{Photoionization Modeling}\label{subsec:xstar}
In order to provide a physical characterization of the highly ionized absorption spectrum, we briefly explored fits using different grids of photoionization models created with \textsc{xstar} \citep[][]{kallman2001}. The fact that we observe He-like and H-like absorption lines requires a high ionization parameter. The prominence of putative $\beta$-lines combined with weakness (absence) of $\alpha$-lines implies a high column density and saturation. 

The best example of dense absorption in the Fe-K band was observed in the black hole LMXB \gro\ \citep[][]{miller2006,miller2008}. The continuum seen in the \chan/HETGS spectrum of that source (an even mixture of a $kT = 1.35$ keV disk and a $\Gamma = 3.5$ power law at an overall luminosity of $L_{\mathrm{X}}\simeq5\times 10^{37}~\lum$) is comparable to the time-averaged spectrum of the X-ray burst in \source\ (which can be roughly described by a $kT_{\mathrm{bb}} \simeq1.8$~keV black body and a $\Gamma \simeq3.7$ power law, yielding $L_{\mathrm{X}}\simeq3.5\times 10^{37}~\lum$).  
The grid used in models 1C and 3C for \gro\ in \citet{miller2008} gives a good fit to the Fe-K absorption spectrum in the X-ray burst of \source\ (Figure~\ref{fig:xstar}). From this fit we measure a column density of $N = (5.0\pm 1.0) \times 10^{23}~\nh$, an ionization parameter of ${\rm log}(\xi) = 5.0\pm 0.5$, and a red shift of $(6\pm 1) \times 10^{3}~ {\rm km}~ {\rm s}^{-1}$. 

Given that $\xi = L / nr^{2}~\ioniz$, we can estimate the distance of the scattering medium from the neutron star. If we assume an X-ray burst luminosity of $L = 10^{38}~\lum$ and that the column density is related to the matter density via $N = nr$, then a radius of $r \simeq (0.3-2.5) \times 10^{3}$~km is implied by our fit. 


\subsection{Energetics of the X-ray Burst}\label{subsec:energetics}
Extrapolating the linear fit to the X-ray burst light curve (Section~\ref{subsec:lc}) down to the intensity of the persistent emission measured in 2008 ($\simeq$5$~\cnts$), suggests that the event had a total duration of $\simeq 1100$~s \citep[$\simeq18$~min; see also][]{degenaar2012_igrburster}. This indicates that the XRT observation covered almost the entire burst tail and provides a good measure of the radiated energy output. 

We integrate the decay fit over the full XRT observation, and apply the count rate to (thermal) bolometric flux conversion factor inferred from time-resolved spectroscopy of the X-ray burst tail (Section~\ref{subsubsec:timeresolved}). This yields a fluence of $f_{\mathrm{XRT}} \simeq 1.1 \times 10^{-5}~\fluence$. 
Combined with the X-ray burst peak (Section~\ref{subsec:bat}), we obtain a total fluence of $f_{\mathrm{b}} \simeq 1.9 \times 10^{-5}~\fluence$. For a source distance of $D=5.0$~kpc, the total radiated energy is $E_{\mathrm{b}} \simeq 6 \times 10^{40}$~erg.


\section{Discussion}\label{sec:discussion}
We investigated an X-ray burst detected from the neutron star LMXB \source\ with \swift\ on 2012 June 25. The duration ($\simeq$18~min) and total radiated energy output ($E_{\mathrm{b}}\simeq5\times10^{40}$~erg) classify it as a rare, energetic intermediate-duration X-ray burst \citep[e.g.,][]{zand05_ucxb,degenaar2010_burst,degenaar2011_burst}. These represent only a few percent of the total number of observed X-ray bursts \citep[][]{falanga08,keek2010}.

During an interval of $\simeq$390--980~s after the X-ray burst peak, the intensity strongly fluctuated by a factor of $\simeq$3 above and below the underlying decay trend, on a time scale of seconds. Out of the thousands of X-ray bursts that have been observed from $\simeq$100 LMXBs \citep[e.g.,][]{galloway06,keek2010}, similar variability has only been reported for five X-ray bursts from four different sources \citep[][]{vanparadijs1990,strohmayer2002,molkov2005,zand05_ucxb,zand2011}. Comparison of their properties suggests that superexpansion is a possible requisite for the occurrence of fluctuations \citep[][]{zand2011}. 
The broad BAT peak, the long XRT tail, and the high radiative energy output all make it plausible that the X-ray burst of \source\ was powerful enough to drive a superexpansion phase. 

Time-resolved (continuum) spectroscopy of superexpansion bursts has shown that the photosphere can be driven out to radial distances of $\simeq10^{3}$~km \citep[e.g.,][]{zand2010,zand2011}.
It has been proposed that superexpansion can disrupt the accretion disk, and that the observed intensity fluctuations are caused by swept up clouds of plasma or puffed-up structures in the disk \citep[][]{zand2011}. If we assume that this plasma moves in Keplerian orbits, then the time scale of the fluctuations seen for \source\ ($P$$\simeq$1--10~s) implies a radial distance of $r=(P^2GM/4\pi^2)^{1/3}\simeq(2-8)\times10^{3}$~km (for $M=1.4~\Msun$). 

Spectral analysis of the X-ray burst tail revealed a highly significant emission line around $\simeq$1~keV in the Fe-L band. This emission feature can be explained through the irradiation of relatively cold gas. If we assume that the line is dynamically broadened, we can estimate the radial distance of this material. The measured width of $\sigma=6.73\times10^{-2}$~keV gives a full width at half maximum of $0.158$~keV. Dividing this by the line energy ($E_{\mathrm{l}} = 1.018$~keV) suggests a velocity of $v\simeq0.16c$. Assuming that the material is in a Keplerian orbit would imply a radial distance of $r=GM/v^2\simeq8\times10^{2}$~km from the neutron star. 

We also found significant absorption features in the Fe-K band; absorption lines at $\simeq$7.73 and $\simeq$8.10~keV, which likely represent Fe \textsc{xxv} He-$\beta$ (7.88 keV) and Fe \textsc{xxvi} Ly-$\beta$ (8.25 keV), as well as K edges consistent with Fe \textsc{xxv} (8.8 keV) and Fe \textsc{xxvi} (9.3 keV). By fitting the absorption features with photoionization models we inferred a radial distance of $r\simeq(0.3-2.5)\times10^{3}$~km for the hot, ionized material (Section~\ref{subsec:xstar}). The spectral fits suggest a modest redshift for the putative absorption lines of $\simeq6\times10^{3}~\mathrm{km}~\mathrm{s}^{-1}$. This could be in line with the radial distance, since the free-fall velocity at $10^3$~km is on the order of $\simeq10^{4}~\mathrm{km}~\mathrm{s}^{-1}$. However, velocity shifts of $\simeq$20--50 eV might also arise through a drift in the gain calibration of the CCD (A. Beardmore, private communication).

Disruption of the accretion disk by super-Eddington fluxes can plausibly have given rise to the Fe-L and Fe-K band atomic features \citep[e.g.,][]{day1991,ballantyne2004,ballantyne2005}. The radial distance found from fitting the Fe-K band absorption features is broadly consistent with that implied by spectral fits to the Fe-L emission line. The latter could arise due to the illumination of cold blobs of gas, which may be pressure-confined within hotter gas that is seen in absorption. Both components of the gas must have rotational support and be executing nearly Keplerian orbits, since the fallback time from a radial distance of $r \simeq 10^{3}$~km is $\simeq$0.1~s. {The distance and velocity estimates arising from our analysis of these atomic features are similar to those inferred from time-resolved continuum spectroscopy of superexpansion X-ray bursts \citep[][]{zand2010}.

We note that Fe-K emission lines and absorption edges have been reported for a handful of X-ray bursts \citep[][]{vanparadijs1990,strohmayer2002,zand2010}. However, to our knowledge it is the first time that such features are unambiguously detected during an X-ray burst at CCD resolution. Fe-L emission has not been seen before; the far majority of X-ray bursts have been observed with instruments that did not have coverage below 2 keV (e.g., \beppo/WFC, \rxte, and \inte). 

The time scale of the strong variations in the X-ray burst light curve imply a similar radial distance as inferred from the spectral absorption and emission features. It is therefore likely that all are caused by the same material and mechanism. We conclude that there are three independent lines of evidence that suggest that the energetic X-ray burst from \source\ involved a superexpansion phase that may have disrupted the (inner) accretion disk out to $\simeq$$10^{3}$~km ($\simeq$50$~R_{\mathrm{g}}$ with $R_{\mathrm{g}}=RM/c^2$ the neutron star gravitational radius).


\acknowledgments
N.D. is supported by NASA through Hubble Postdoctoral Fellowship grant number HST-HF-51287.01-A from the Space Telescope Science Institute. R.W. acknowledges support from a European Research Council starting grant. This work made use of the \swift\ public data archive.

{\it Facility:} \facility{Swift (BAT,XRT)}

\end{document}